\documentclass[conference]{IEEEtran}
\IEEEoverridecommandlockouts

\usepackage[T1]{fontenc}
\usepackage[english]{babel}
\usepackage{blindtext}
\usepackage{tikz}
\usepackage{amsmath}
\usepackage{tabu}
\usepackage{adjustbox}
\usepackage{filecontents}
\usepackage{xspace}
\usepackage{titlesec}
\usepackage[font=small]{caption} 

\titlespacing*{\section}
{0pt}{1.5ex}{2.0ex}

\definecolor{darkerblue}{RGB}{0,63,202}
\definecolor{sabzseyedi}{RGB}{14,109,25}
\definecolor{LightGray}{RGB}{247,247,247}

\newcommand{\Authority}{Producer\xspace}
\newcommand{\authority}{producer\xspace}
\newcommand{\hub}{connector\xspace}
\newcommand{\Hub}{Connector\xspace}
\newcommand{\authoritys}{producers\xspace}
\newcommand{\hubs}{connectors\xspace}

\newcommand{\algo}{HackerScope\xspace}
\newcommand{\hscore}{HackerScore\xspace}   
   
\newcommand{\PHS}{PHS\xspace} 
\newcommand{\CHS}{CHS\xspace} 
\newcommand{\PHSlong}{\Authority HackerScore\xspace}
\newcommand{\CHSlong}{\Hub HackerScore\xspace}

\newcommand{\github}{GitHub\xspace}

\newcommand{\malauth}{7389\xspace} 

\newcommand{\malauthor}{malware authors\xspace}
\newcommand{\MalAuthors}{Malware authors\xspace}
\newcommand{\malHackersNo}{30\xspace}

\definecolor{dgreen}{RGB}{30,100,80}
\definecolor{dred}{RGB}{150,0,10}

\definecolor{dpurple}{RGB}{100,30,80}

\usepackage{fancyhdr}
\usepackage{kantlipsum}
\fancypagestyle{plain}{
\fancyhf{}
\fancyhead[C]{\textbf{2020 IEEE/ACM International Conference on Advances in Social
Networks Analysis and Mining (ASONAM)}} 

}
\usepackage{eso-pic}

\begin{document}

\AddToShipoutPictureBG*{
\AtPageUpperLeft{
\setlength\unitlength{1in}
\hspace*{\dimexpr0.5\paperwidth\relax}
\makebox(0,-0.75)[c]{\textbf{2020 IEEE/ACM International Conference on Advances in Social
Networks Analysis and Mining (ASONAM)}}}}



\title{\algo: The Dynamics of a Massive Hacker Online Ecosystem}

\author{
{\rm Risul Islam}\\
UC Riverside\\
risla002@ucr.edu
\and
{\rm Md Omar Faruk Rokon}\\
UC Riverside\\
mroko001@ucr.edu
\and
{\rm Ahmad Darki}\\
UC Riverside\\
adark001@ucr.edu
 \and 
{\rm Michalis Faloutsos}\\ 
UC Riverside\\
michalis@cs.ucr.edu
} 

\maketitle

\IEEEoverridecommandlockouts
\IEEEpubid{\parbox{\columnwidth}{\vspace{8pt}
\makebox[\columnwidth][t]{IEEE/ACM ASONAM 2020, December 7-10, 2020}
\makebox[\columnwidth][t]{978-1-7281-1056-1/20/\$31.00~\copyright\space2020 IEEE} \hfill}
\hspace{\columnsep}\makebox[\columnwidth]{}}
\IEEEpubidadjcol

	
\begin{abstract}

Authors of malicious software are not hiding as much as one would assume: they have a visible online footprint.
Apart from online forums, this footprint appears in software development platforms, where authors create publicly-accessible malware repositories to share and collaborate.
With the exception of a few recent efforts, 
the existence and the dynamics of this community has received surprisingly limited attention.
The goal of our work  is to analyze this  ecosystem of hackers in order to: (a) understand their collaborative patterns, and (b) identify and profile its most influential authors. 
We develop \algo, a systematic approach for analyzing the dynamics of this hacker ecosystem.
Leveraging our targeted data collection,
we conduct an extensive study of \malauth authors of malware repositories on \github, which we combine with their activity on four security forums.
From a modeling point of view,
we study the ecosystem using three network representations: (a) the author-author network, 
(b) the author-repository network,
and (c) cross-platform egonets. 
Our analysis leads to the following key observations:
(a) the ecosystem is growing at an accelerating rate as the number of new malware authors per year  triples every 2 years,
(b) it is highly collaborative, more so than the rest of \github authors, and
(c) it includes influential and professional hackers. We find \malHackersNo
authors maintain an online ``brand" across \github and our security forums.
Our study is a significant step towards 
using public online information for understanding
the malicious hacker community.

\end{abstract}


\pagestyle{plain} 

\section{Introduction}
 

{\em ``How can a 17 year old kid from Florida~\cite{17yearsoldboy} be reportedly the mastermind behind the recent hacking of Twitter?
} This question is part of the motivation behind this work.

The security community has a fairly limited understanding of malicious hackers  and their interactions. As a result,
security practitioners do not really know their ``enemy”.
On the one hand, the hacker community is fairly wide encompassing curious teenagers, aspiring hackers, and professional criminals.
On the other hand, the hackers are surprisingly bold in leaving a digital footprint, if one looks at the right places in the Internet. 
 For example, there are various online forums, where
 hackers not only share information, but they also boast of their successes.

How can we begin to understand the  ecosystem of malicious hackers based on their online footprint? The input is the online activities of these hackers, and the
goal is to answer the following questions:
    (a)  do these hackers work in groups or alone,
    and (b) who are the most influential hackers?
Here, we consider two types of platforms  that hackers frequent: (a) software archives, and (b) online security forums.
It turns out that  popular and public software archives, such as \github harbor {\bf malware authors}, who create publicly-accessible 
malware repositories~\cite{rokon2020source}. 
Furthermore, online forums have recently emerged as marketplaces and information hubs of malicious activities~\cite{Joobin, portnoff2017tools}.
In the rest of this paper, we will use the term {\bf hacker} to refer 
to actors who develop and use software of malicious intent. We will also use the term {\em hackers}
and  {\em malware authors} interchangeably, 
although some malware authors may not have malicious intent. 


There is limited work for the problem as defined above.
First, we are not aware of a study that  systematically profiles the dynamics of the online
hacker ecosystem, and especially one considering software archives.
Most of the previous efforts on \github follow a software-centric view  or study \github at large without focusing on malware~\cite{calleja2016look}~\cite{calleja2018malsource}~\cite{blincoe2016understanding}. 
Most of the previous works on online forums focus on identifying emerging topics and threats \cite{Joobin, portnoff2017tools}.
Other efforts report malware activity, focusing on hacking events,
and much less, if at all, on the ecosystem of hackers~\cite{Sapienza2017_USC1, Sapienza2018_USC2}. 
We elaborate on previous works in Section~\ref{sec:related}.

We propose {\bf \algo}, a systematic approach for modeling  the  ecosystem  of  malware authors by analyzing their online footprint.
We start with an extensive analysis of malware authors on \github, as this is a significantly less-studied space.
We then use security forums to find more information about these authors.
From an algorithmic point of view,  we use three network representations: 
(a) the author-author network,
(b) the author-repository network,
and (c) cross-platform egonets, which we explain later. In addition, we use some basic Natural Language Processing techniques, which we intend to develop further in the future.

\begin{figure}[h]
    \centering
    \includegraphics[width=0.85\linewidth, height=4.5cm]{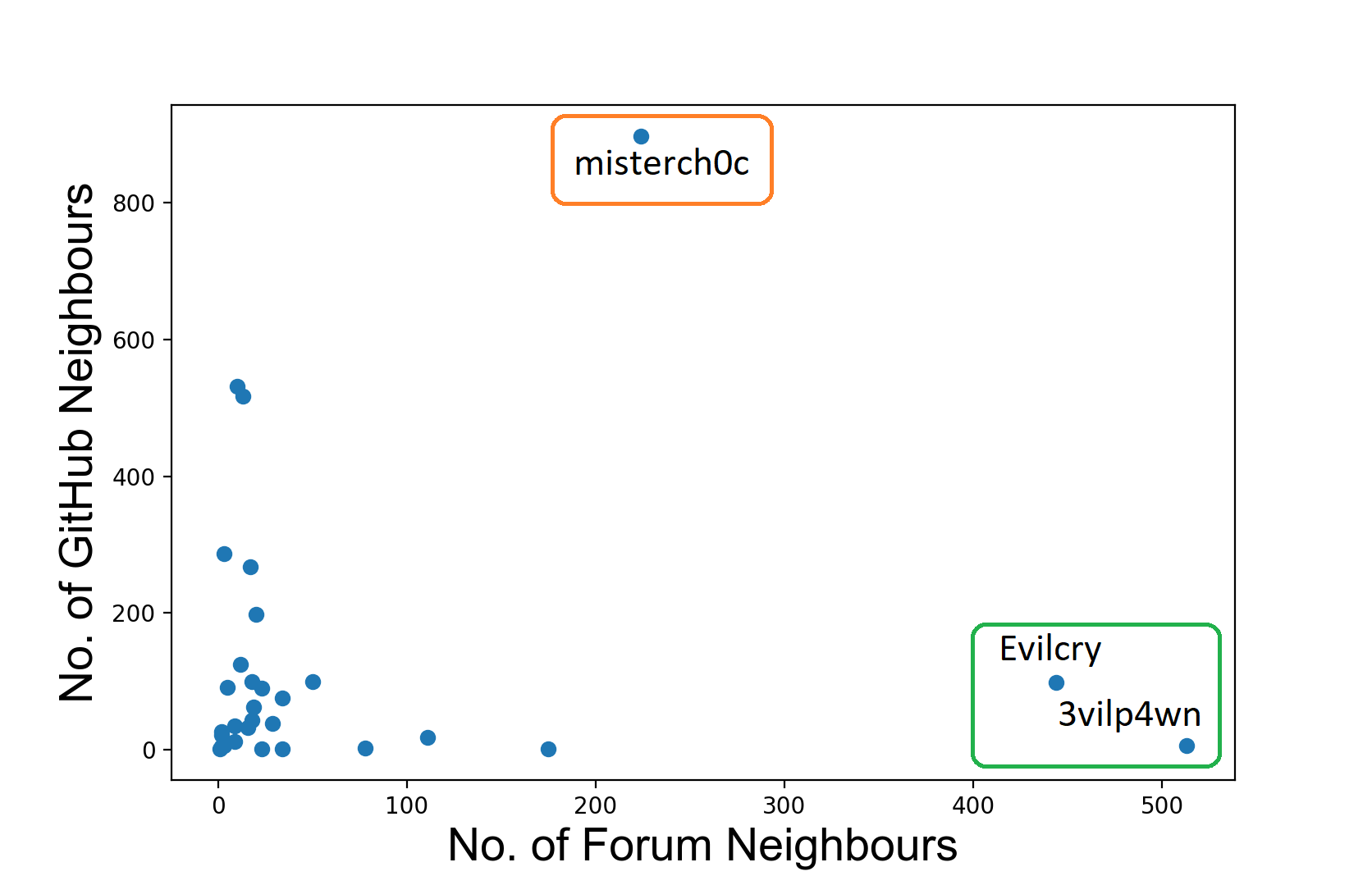}
    \caption{Profiling hackers across platforms using our cross-platform egonet: the scatter-plot of  the number of neighbors on \github  versus  those on security forums for 30 malware authors
    as captured in our cross-platform egonet.
    }
    \label{fig:scree-egonet}
\end{figure}

We apply and evaluate our approach
using \malauth malware authors on \github 
over the span of 11 years and leverage the activity on four security forums
 in the grey area between white-hat and black-hat security.
\github is arguably the largest repository with roughly 30 million public repositories, while, appropriately fine-tuned, our approach can be used on 
other software archives.
Our approach encompasses four research thrusts, which identify and model:
(a) statistics and trends,
(b) communities of hackers and their dynamics,
(c) influential hackers,
and (d) hacker profiles across  different online platforms.
For the latter type, we show the collaborators of hackers
as captured by the cross-platform egonets spanning \github and security forums in Figure~\ref{fig:scree-egonet}.
Our key results are summarized in the following points.

{\bf a. The ecosystem is growing at an accelerating rate:} The number of new malware authors on \github is roughly tripling every two years. This alarming trend points to the importance of monitoring this ecosystem.


{\bf b. The ecosystem is highly collaborative:} 
We find  513 collaboration communities on \github with 
high cohesiveness ({\it Modularity Score} within [0.65-0.78]),
including many large communities with hundreds of users.
The malware community is very collaborative:
 a malware repository is forked {\em four times} more compared to
a regular \github repository. 

{\bf c. We identify a group of 1.7\% of influential authors:}
We develop a systematic approach to determine the influence among malware authors.
Our novelty lies in: (a) considering many types of interactions,
and (b) capturing the network-wide influence of an author.
We find a core group of 1.7\% of the malware authors, who are responsible 
for: (a) generating influential repositories, and (b) providing the social
backbone of the malware community.



{\bf d. We identify professional hackers in the ecosystem:} 
We find that \malHackersNo  authors are professional {\it malicious} hackers.
Going across platforms, we find \github authors  
who are quite active on our security forums.
We show the evidence that these
are professional hackers, who are building 
an online ``brand". 
 For example,  user {\em 3vilp4wn} is the author of a keylogger repository on 
 \github, which he promotes in  
 the {\em HackThisSite} forum using the same username
 (shown at bottom right in Figure~\ref{fig:scree-egonet}).


{\bf Our work in perspective.} 
The proposed work is part of an ambitious goal: we want to model the Internet hacker ecosystem at large as it manifests itself across platforms.
Our initial results are promising: a) the hackers seem to want to establish a brand, hence they want to be visible, and b) a cross-platform study is possible, as some authors maintain the same login name.
Our systematic approach here constitutes a building block towards the ultimate goal. 
With appropriate follow up work, achieving this goal can have a huge
practical impact: security analysts could prepare for emerging threats,  anticipate malicious activity, and identify their perpetrators.




{\em Open-sourcing for maximal impact.}
We use Python v3.6.2 packages to implement all the modules of \algo. We intend to make our datasets and tools public for research purposes.

\section{Background and Data}
\label{sec:data}




Our work focuses on \github, the largest software archive with roughly 30 million public repositories, and uses data from security forums.  Although \github policies do not allow malware, authors do not seem to abide by them.

{\bf A. \github data.} \github platform 
enables software developers to create software repositories in order to  store, share, and collaborate  on projects and provides many social-network-type functions.

We  define some basic terminology here. 
We use the term {\it author} to describe a \github user who has created at least one repository.
A {\it malware repository} contains  malicious software and
a {\it malware author} owns at least one such repository.
Users 
can {\it star, watch} and {\it fork} other {\em malware repositories}. {\it Forking} means creating a clone of another repository. A forked repository is sometimes merged back with the original parent repository, and we call this a {\it contribution}. Users can also {\it comment} by providing suggestions and feedback to other authors' repositories.

We use a dataset of 7389 malware authors 
and their related 8644 malware repositories, which were identified among 97K repositories in our prior work~\cite{rokon2020source}.
This is arguably the largest malware archive of its kind with repositories  spanning roughly 11 years. These repositories have been identified as malicious with a very high precision 
(89\%). 
Note that the queries with the \github API, which were used in the data collection,  return primary or non-forked repositories.
A discussion on the process, accuracy, and validity of the dataset can be found in the original study~\cite{rokon2020source}.




For each malware author in our dataset, we have the following information: (a) the list of the malware repositories created by her,  and (b) the list of followers.
For each malware repository, we have the lists of users, who:
(a) star, (b) watch, (c) fork, (d) comment, or (e) contribute to the repository. 

{\bf Repository metadata.} Each repository is also associated with
a set of user generated fields, such as title, readme file, description.  We can use this {\em metadata} to extract information about the repository. 
We leverage our earlier work where we discuss the processing
of this metadata in more detail~\cite{rokon2020source}.

For a given repository, a security expert would want to know: (a)
the type of malware (e.g. ransomware and keylogger), and (b) the target platform (e.g. Linux and Windows).
For this, we define  two  sets of keywords: (a) 13 types of malware, $S_1$
and (b) 6 types of target platforms, and $S_2$. 
Figure~\ref{fig:worldcloud_community} provides a visual list of these two sets of keywords.
We define the Repository Keyword Set, {\em $W_r$}, for repository $r$, as a set consisting of the keyword sets $S_1$ and $S_2$
that are present in its metadata.
Clearly, one can extend and refine these keyword sets,
to provide additional information, such as the programming language in use, which we will consider in the future.
Note that our earlier work provides evidence
that using this metadata as we do here can provide fairly accurate and useful information~\cite{rokon2020source}.



\begin{table}[t]
    \caption{Our four online security forums. 
    }
    \footnotesize
    \centering
    \begin{tabular}{|p{0.27\linewidth}|p{0.16\linewidth}|p{0.16\linewidth}|p{0.18\linewidth}|}
         \hline
         \textbf{Forum} & \textbf{Users} & \textbf{Threads} &  \textbf{Posts}\\
         \hline
         Offensive Comm. & 5412 & 3214 & 23918 \\
         \hline
         Ethical Hacker &  5482 & 3290 & 22434 \\
         \hline
         Hack This Site & 2970 & 2740 & 20116 \\
         \hline
         Wilders Security & 3343 & 3741 & 15121 \\
         \hline
    \end{tabular}
    \label{tab:statforums}
\end{table}

{\bf B. Security forum data.} We also utilize  data that we collect from four security forums: Wilders Security, Offensive Community, Hack This Site, and Ethical Hackers ~\cite{secforums}. In these forums,
users initiate discussion threads 
in which other interested users can post to share their opinion.
Each tuple in our dataset contains the following information: forum ID, thread ID, post ID, username, and post content. We provide a brief description of our forums below, and an overview of key numbers in Table~\ref{tab:statforums}.

{\bf a.  OffensiveCommunity (OC):} As the name suggests, this forum contains ``offensive security” related threads, namely, breaking into systems. Many posts consist of step by step instructions on how to compromise systems, and advertise hacking tools and services.

{\bf b. HackThisSite (HTS):} As the name suggests, this forum has also an attacking orientation. There are threads that explain how to break into websites and systems, but there are also more general discussions on cyber-security.

{\bf c. EthicalHackers  (EH):} This  forum  seems  to   consist mostly of ``white-hat" hackers, as its name suggests. However, there are many threads with malicious intentions in this forum.

{\bf d. WildersSecurity (WS):} The threads in this forum fall in the grey area, discussing both ``black-hat" and ``white-hat" skills.

\section{Our Approach} 
\label{sec:approach}

\begin{figure}[t]
    \centering
    \includegraphics[width=0.95\linewidth]{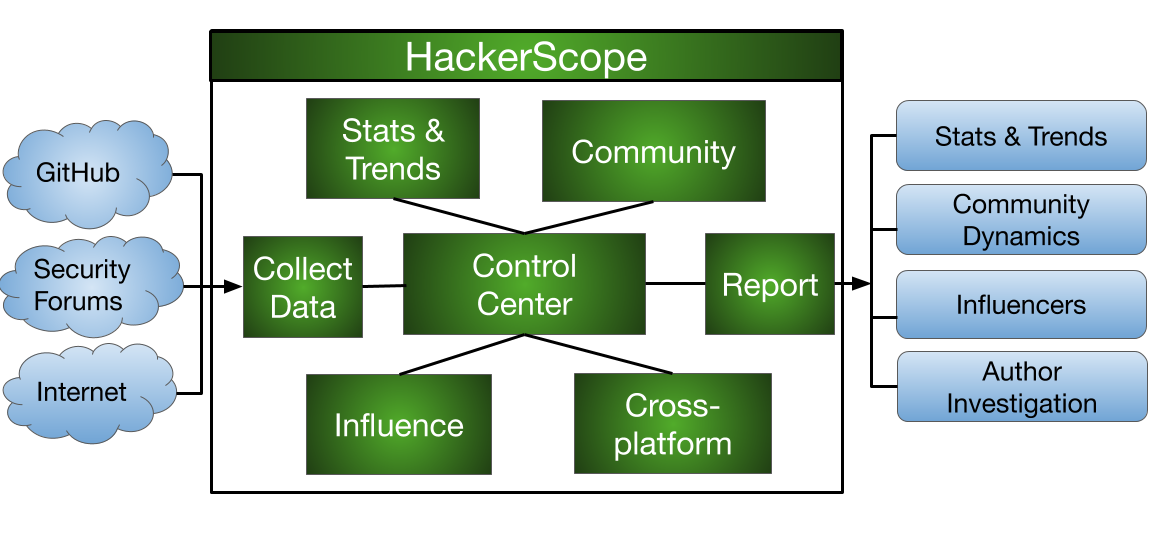}
    \caption{The overview of our approach highlighting the key functions.
    }
    \label{fig:overview}
\end{figure}

We have an ambitious vision for our approach,
which we plan to release as a software platform.
We provide a brief overview in Figure~ \ref{fig:overview}.
In this paper, we will elaborate on the four analysis modules:
(a) a statistics and trends module, which provides the landscape of primary behaviors of the ecosystem (Section \ref{sec:basic_temporal}),
(b) a community analysis module, which identifies and profiles communities of collaboration
(Section \ref{sec:collaboration}),
(c) an influence analysis module, which defines and calculates the significance of authors
(Section \ref{sec:influence}),
and (d) cross-platform analysis module (Section \ref{sec:intention}).

In addition, our approach also includes: 
 a data collection  module, which aggregates, cleans and preprocesses the raw information;
 a  control center module; and
 a reporting module.
These modules are not equally developed, while at the same time, we could not provide all the types of results that we have available due to space limitations.

Below, we highlight some interesting or novel aspects of our approach, which are often cutting across several modules.

{\bf a. Synthesizing multi-source data.} Our approach focuses on data 
for authors from \github and combines it with additional data from security forums, and
Internet searches.

 {\bf b. Defining appropriate features.} As we already saw, the authors and the repositories have a very rich set of interactions. We have primary (measured directly) and secondary (derived from the primary) features, which need to be determined carefully to capture effectively the dynamics of the ecosystem. These interactions go beyond a simple ``friend" relationship of other social media.

 {\bf c. Modeling the dynamics.} We use three network representations to capture the rich interactions  and  relationships among authors and repositories. The network representations include: (a) the author-author network, 
 (b) the author-repository network,
 and (c) cross-platform egonets. 

 {\bf d. Reporting behaviors.} The goal is to provide intuitive and actionable information
 in an appealing and ideally interactive fashion.
 The  results in this paper provide an indication of
 some initial
 plots and tables that our approach will provide to the end user,
 who could be a researcher or a security analyst.

\section{Statistics and Trends}
\label{sec:basic_temporal}

This section describes the functionality of the {\em statistics 
and trends} module of our approach, whose intention is to provide
a basic understanding of author behaviors.




\textbf{A. Basic distributions of malware authors.}
We study the  complementary 
cumulative distribution function (CCDF) of three metrics: 
(a) the number of repositories created, (b) the number of followers, and (c) sum of the number of forks across all the malware repositories of the author. 
As expected all distributions are skewed, but
the plots are omitted due to space constraints.
First, we find that 15 authors are contributing roughly 5\% of all malware repositories, while 99\% of all authors have created less than 5 repositories each. Second, we find that 3\% (221) of the authors  have more than 300 followers each, while  70\% of the authors have less than 16 followers. Finally, examining the total number of forks per author, we find that 3\% (221) of the authors  have their repositories forked more than 150, while 43\% of authors encounter at least one fork.

\textbf{B. Forking behavior: Malware repositories are forked four times more than the average repository.}
Malware repositories are more aggressively forked, which
is an indication of the higher collaboration in the ecosystem.
First, we find that
a malware repository is forked 4.01 times on average, while
a regular \github repository is forked 0.9 times, as reported
in previous studies~\cite{jiang2017and}.
Second, we want to see if this is due to a few popular repositories, but this is not the case. We find that 39\% of the malware repositories are forked at least once, 
while this is true for only 14\% for general repositories ~\cite{jiang2017and}. 

\begin{figure}[t]
    \centering
    \includegraphics[width=0.85\linewidth, height=3.5cm]{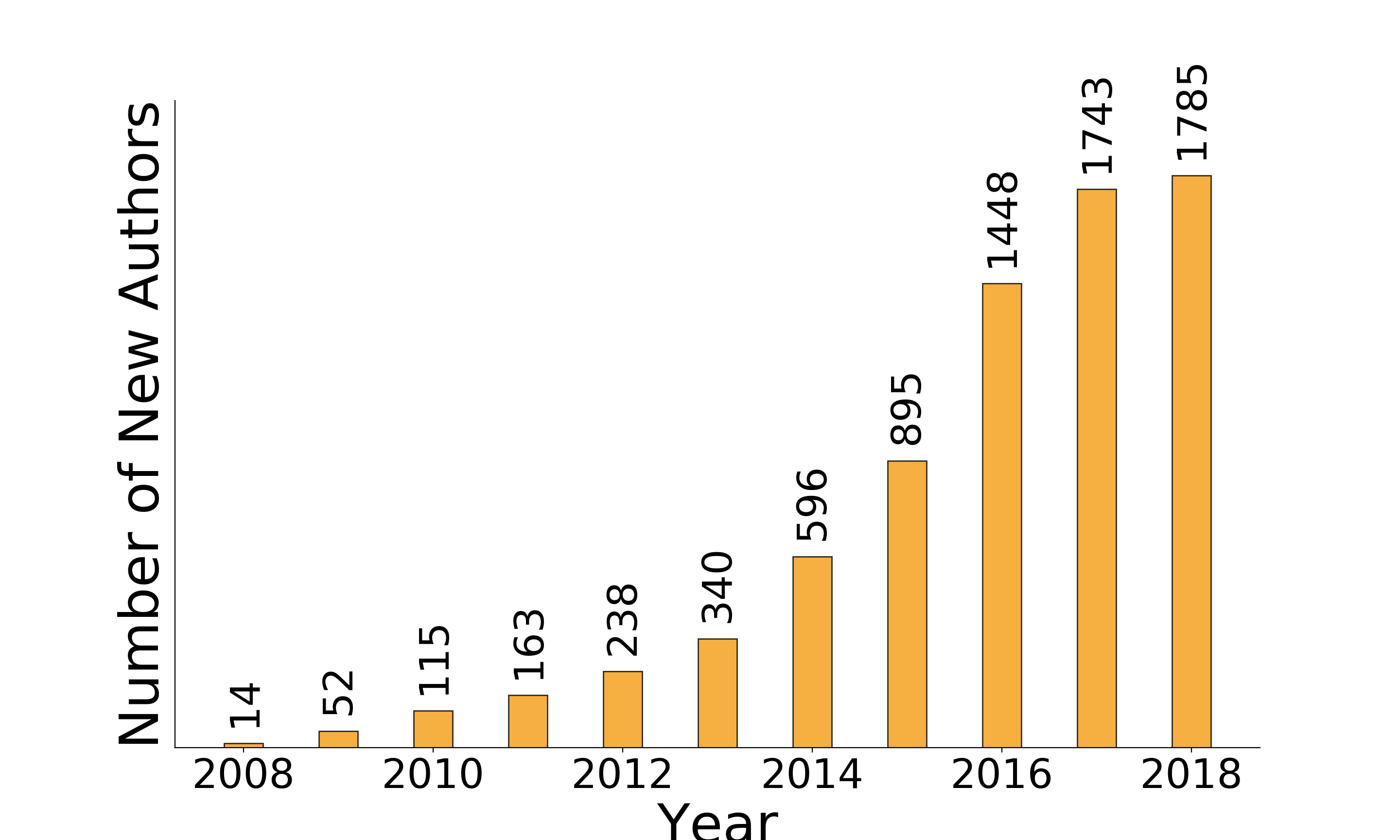}
    \caption{New malware authors in the ecosystem per year.
    }
    \label{fig:trend_author}
\end{figure}

\textbf{C. Trends.}
{\it ``How fast is this ecosystem growing?"}
To answer the question,  we plot the number of new malware authors per year in Figure~\ref{fig:trend_author}. 
We consider that an author joins the ecosystem at the time that
they create their first malware repository in our database. 


{\bf  a. The number of new malware authors almost triples every two years.}
  We plot the new malware authors per year in Figure~\ref{fig:trend_author}.
We observe an increase from 238 malware authors in 2012 to 596 authors in 2014  and to 1448 authors in 2016.
We also observe a steep 62\%  increase from 2015 to 2016. This trend is interesting and alarming at the same time.

{\bf b.  The number of new malware repositories  more than triples every four years.} 
Echoing the growth of the authors, the number of repositories is also increasing super-linearly. 
In the future, we plan to study the trends of malware in terms of
both types of malware and its target platform.

\section{Identifying Influential Authors}
\label{sec:influence}


To understand the dynamics of the ecosystem,
we want to answer the following question:
{\it ``Who are the most influential authors?''}
The functionality in this section is part of the {\em influence analysis} module of Figure~\ref{fig:overview}.



{\bf A. \hscore: Identifying influential authors.} 
We argue that finding influential authors presents several challenges.
First, there are many different activities and interactions, such as creating repositories, commenting, following other authors and being followed by other authors.
Second, we can consider two  types of  actions: (a) creating influential artifacts,  (b) observing and engaging with other people and artifacts.
Furthermore, the distinction is not always clear. For example, forking a repository
creates a new, but derivative, repository.

To address the above challenges, we take  {\bf socially-aware approach} to influence: creating
a few influential repositories is more important than creating many non-influential repositories. We discuss how we model and calculate this influence below.

{\bf The Author-Author graph (AA).}
We create the  Author-Author network to capture the
network-wide interaction among authors.
We define a weighted labeled multi-digraph: $G(V, E, W, L_e)$ where $V$ is the malware author set,  $E$ is the set of edges, $W$ is the weight set and $L_e$ is the set of labels that an edge $e$ can be associated with. These labels correspond to different types of relationships between authors. Here we opted to consider only malware authors in the graph to raise the bar for being
part of the hacker community. 

{\bf  The types of interactions.} 
We consider four types of relationships between authors here.
A directed edge
$(u,v)$ from author $u$ to $v$ can be (i) a follower edge: when $u$ follows $v$, (ii) a fork edge: when $u$ forks a repository of $v$, (iii) a contribution edge: $u$ contributes code in a repository of $v$, and (iv) a comment edge: $u$ comments in a repository of $v$. 
These relationships capture the most substantial author-level interactions.





{\bf The multi-graph challenge and weight calibration.} Our graph consists
of different types of edges, which represent different relationships that 
we want to consider in tandem.
The challenge is that the relationships have significantly different distributions,
which can give an unfair advantage or eliminate the importance of a relationship.
For example, contribution activities are rarer compared to following, but one can argue that a contribution to a repository is a more meaningful relationship and it should be given appropriate weight.

For fairness, we make the weight of a type of edge inversely proportional to 
a measure of its relative frequency. In detail,
we calculate the average degree $d_{type}$ for each {\em type} of edge: follower, fork, contribution, and comment from the subgraph containing only that type of edges from the AA graph. We find the following average degrees: $d_{follower} = 12.21$, $d_{fork} = 4.67$, $d_{contribution} = 0.53$ and $d_{comment} = 0.49$. 
We normalize these average degrees using the minimum average degree ($d_{min} = 0.49$) and we get the {\em inverse} of this  value as the weight for that edge, namely, $d_{min}/d_{type}$. 
This way, we set the following weights:
 $w = 0.04$ for a following edge,  $w = 0.1$ for a forking edge, 
and  $w = 1$ for a commenting or a contribution edge. 
This enables us to consider each relationship type more fairly and meaningfully.



We propose a socially-aware and integrated approach to combine all the 
author activities in a single framework.
First, we identify and define two roles in the ecosystem: 
(a) {\bf \authoritys}, who create influential malware repositories,
and (b) {\bf \hubs}, who enhance the community 
by engaging with influential malware authors and repositories. 
To calculate the roles of the malware authors, we first model the interaction among authors in the {\bf AA} graph described above. Next, we  apply our algorithm, a customized version
of a weighted hyperlink-induced topic search  algorithm 
modified to handle the multiple types of relationships between authors.
We discuss the related algorithms in Section~\ref{sec:related}.


{\bf Calculating the \hscore.}
We associate each node $u$  with two values: (a)  a {\bf \PHSlong} value, $\PHS_u$, and 
(b) {\bf \CHSlong} value, 
$\CHS_u$. 
Let $w(u,v)$ be the weight of edge $(u,v)$ based on its label, as discussed above.

The algorithm  iterative refines the \authority and \hub values until it converges. We, now, elaborate on the steps.
First,  $\PHS_u$ and $\CHS_u$ are initialized to 1.
During the iterative step,
the algorithm updates the values as follows: (i) for all $v$ pointing to $u$: $\PHS_u = \sum_{v} w(v,u)*\CHS_{v}$, or zero in the absence of such edges, (ii) for all $z$ pointed by $u$: $\CHS_u = \sum_{z} w(u,z)*\PHS_{z}$, or zero in the absence of such edges, and (iii) we normalize $\PHS_u$ and $\CHS_u$, so that $\sum_{u}\PHS_u = \sum_{u}\CHS_{u} = 1$.
For the convergence, we set a tolerance threshold of $10^{-9}$ for the change of the value of any node. After 449 iterations,
we obtain the two \hscore values for each author.

\begin{figure}[t]
    \centering
    \includegraphics[width=0.85\linewidth,height=4cm]{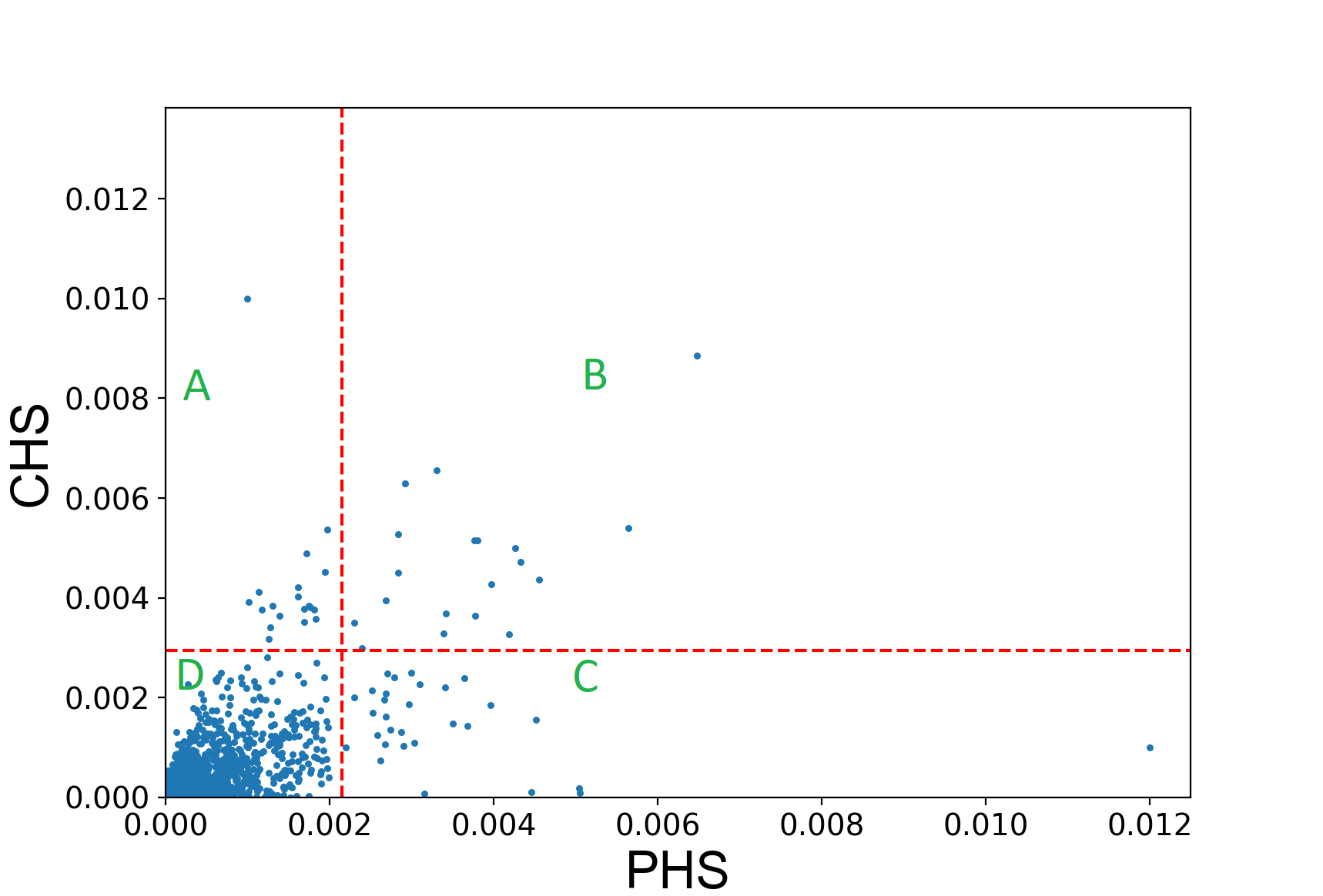}
    \caption{The scatterplot of the \CHSlong vs. \PHSlong for the malware authors in our \github dataset. 
    }
    \label{fig:chsvsphs}
\end{figure}

{\bf Identifying influential malware authors.} In Figure~\ref{fig:chsvsphs},
we plot the \CHSlong versus \PHSlong for our malware authors.
Separately, we identify ``knees" in the individual distributions of each score
at \PHS= 0.00215 and \CHS= 0.0029 indicated by the red dotted lines.
 This way, we observe four regions defined by the combination of low and high values for  \PHS and \CHS values
 which shows if an author is influential as \authority or \hub.
 
 {\bf A few authors (1.7\%) drive the community.}
The three regions of influence together consist of 128 malware authors (1.7\%).
The break down of the region size is fairly even:
Region A  of mostly \hub authors devoted to connect the malware community is 0.6\%, Region C of the influential \authoritys who are the originator of the malware resources is 0.7\%, and Region B of dual influence is 0.4\%.
We use the term {\bf Highly Influential Group (HIG)} to refer to this group of authors.

We provide a profile overview of the two most influential authors per region in Table~\ref{tab:top7authors}.
 The most influential author of Region C is {\it cyberthrets}, with the highest \PHS (0.012) and 336 malware repositories. She gained a huge following by creating all her repositories of assembly code malware on Feb 16, 2016.
The top \hub author from Region A is {\it critics}
with a \CHS score of 0.01, which stems from her 446 comments across 301 repositories.
The top malware author from Region B is {\it D4Vince} for his dual role in producing credential reuse tools with 7 repositories and 
165 comments and 187 contributions.

{\bf The importance of socially-aware significance.} We argue that our socially-aware definition of
significance provides more meaningful results  than simply taking the top-ranked users in any primary metric in isolation. 
First, the two scores capture different aspects of influence: they can differ by orders of magnitude as is the case with {\it cyberthrets} and {\em ytisf}.
Second, our scores capture a combined network-wide influence that each 
primary metric could miss. For example,
our most influential \authoritys do not always own  many malware repositories. Malware author {\it D4vince} and {\it n1nj4sec}, mentioned in Table \ref{tab:top7authors}, have single-digit repositories (7 and 8 respectively) and yet are two of the top \authoritys. On the other hand, author {\it kaist-is521} is ranked way below than {\it n1nj4sec} in terms of \hscore (\PHS=0.0001 and \CHS=0.00013), although she has 18 malware repositories. 

\begin{table}[t]
    \centering
    \footnotesize
    \caption{The profiles of the two most influential malware authors from each region A, B, and C.  
    }
    \begin{tabular}
        {|p{0.12\linewidth}|p{0.07\linewidth}|p{0.07\linewidth}|p{0.07\linewidth}|p{0.07\linewidth}|p{0.07\linewidth}|p{0.08\linewidth}|p{0.063\linewidth}|}
        \hline
         \textbf{Name} & \textbf{\PHS} & \textbf{\CHS} &\textbf{Repos}& \textbf{Follow -ers}& \textbf{Forks}& \textbf{Com- ments }&\textbf{Cont-rib/s}\\ \hline
         cyberthrets & {\bf 0.012} & 0.001 & 336& 1013 & 778 & 13 & 2\\ \hline
         ytisf & {\bf 0.005} & $10^{-6}$ &12& 606 & 1412 &  4 & 1\\ \hline
         
         critics & 0.001 & {\bf 0.01} & 6& 396 & 83 & 446& 301 \\ \hline
         samyk & 0.0018 & {\bf 0.0058} &2& 554 & 125 & 176& 209\\ \hline
         
         D4Vince & {\bf 0.0066} & {\bf 0.0082} & 7 &  608 & 499 & 165 & 187\\ \hline
         n1nj4sec & {\bf 0.0058} & {\bf 0.0052} & 8& 876 & 1391 & 64& 79\\ 
         \hline
    \end{tabular}
    \label{tab:top7authors}
\end{table}

{\bf B. Reciprocity of interactions.}
We want to understand better the nature
of the author interactions here.

{\it ``Is the influence among malware authors reciprocal?''} 
The answer is negative: {\bf the relationships are not reciprocal}, which is in stark contrast to the reciprocal relationships
in other social media like Twitter and Facebook \cite{weng2010twitterrank}.
We consider a total of six relationships: following, forking, commenting, contributing, watching, and starring relationships. We define the {\bf Reciprocity Index}
 for relationship {\em x}, $RI_{x}$, to be  the ratio of reciprocal relationships over the pairs of authors with that type of relationship (unilateral or mutual) in the Author-Author network.

We find that the reciprocity is low and less than 7.3\% for all the relationships in question. 
 By contrast, reciprocity  is often above 70\%  in social media, like Facebook or Twitter~\cite{weng2010twitterrank}. These 
 social media mirror personal relationships  and have an etiquette of conduct. 
 We conjecture that the lower reciprocity on \github 
 is due to its  utilitarian orientation:
 following an author 
 stems from a professional interest.

\section{ Community Analysis}
\label{sec:collaboration}


This section describes the functionality of the {\em community analysis} module,
whose goal is to 
identify the communities of collaboration among the malware authors on \github.

{\bf A. Identifying  collaboration communities.}
We quantify the collaborative nature of the  malware authors as follows. 

{\bf The Author-Repository graph (AR).} 
We define the Author-Repository graph to be an undirected bipartite graph, $G=(A,R,E)$, where $A$ is the set of malware authors and $R$ is the set of malware repositories. 
An edge $(u,r) \in E$ exists, if  author $u$: (a) creates, (b) stars, (c) forks, (d) watches, (e) comments, or (f) contributes to repository $r$.



{\bf Identifying bipartite communities.} To identify  communities, we employ a 
greedy modularity maximization 
algorithm modified for bipartite graphs as we discuss in our related work.

\begin{figure}[t]
    \centering
    \includegraphics[width=0.8\linewidth,height=4cm]{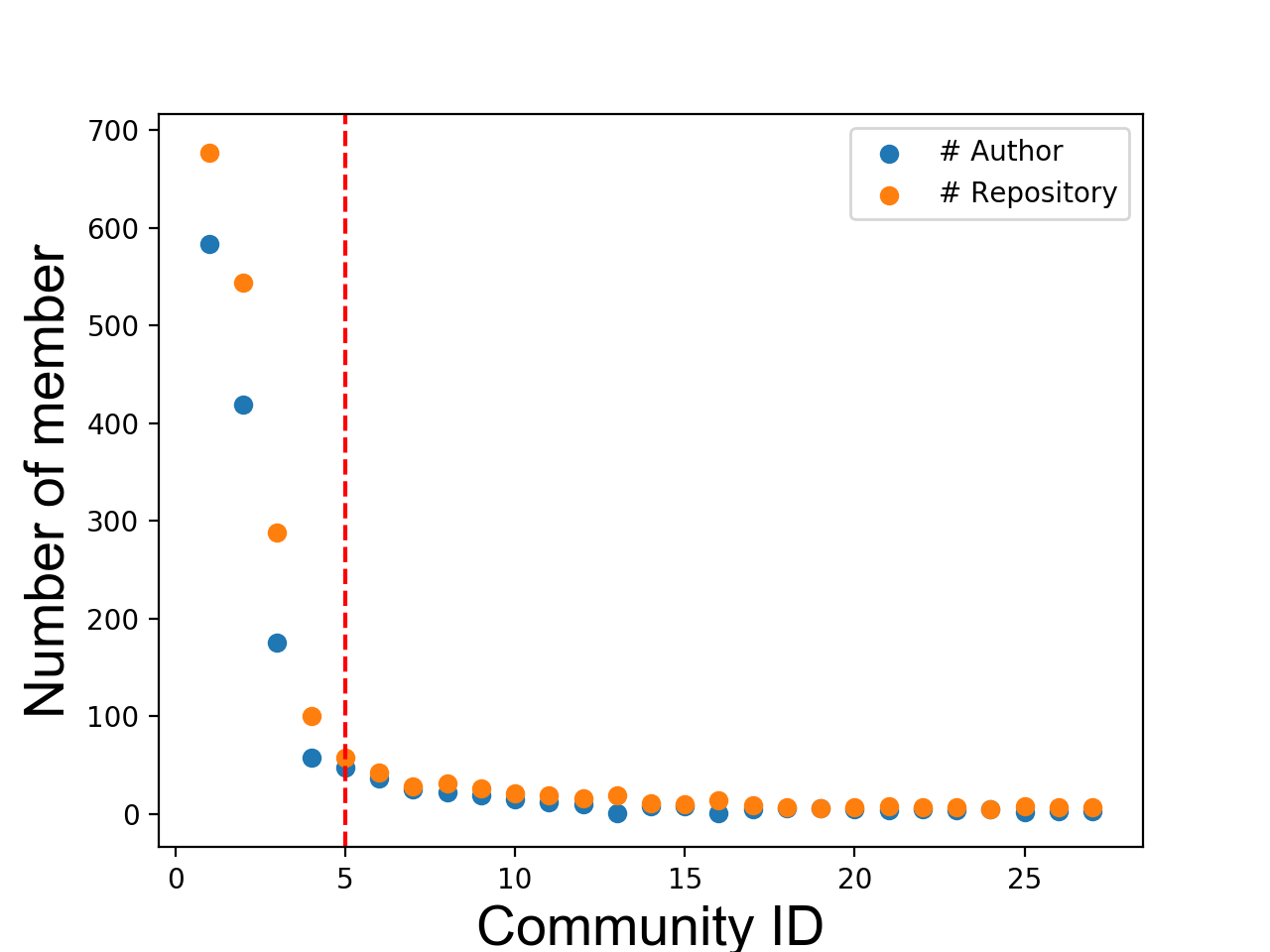}
    \caption{The distribution of the number of authors and repositories for the 27 largest communities in the order of community size.
    }
    \label{fig:knee_community}
\end{figure}

We find a total of 513 communities 
spanning a wide range of sizes 
as shown in Figure~\ref{fig:knee_community}.
The size of the communities follows skewed distribution. 
In Figure ~\ref{fig:knee_community}, we plot the number of \malauthor and repositories per community in order of decreasing community size (defined as the sum of authors and repositories).
We find that 90\% of communities have less than 14 authors and repositories.
We also see a fairly sharp knee in the plot at the fifth  community,
as shown by the vertical line.

\textbf{B. Profiling the communities.}
A full investigation of the  purpose, evolution, and internal structure of each community  could be a research topic in its own right. 
Here, we only provide 
an initial investigation around the following three questions.



{\it a. How cohesive are our communities?} We report the {\bf Modularity Score \textit{($MS_C$)}}, which quantifies
 the cohesiveness of a community $C$.
The $MS_C$ is defined as follows: $MS_C = \frac{n_C(E)}{N_C(E)}$, where $n_C(E)$ is the total number of edges and $N_C(E)$ is the number of all possible edges in community $C$ (if the community was a bipartite clique). 

Overall, {\bf our communities are highly cohesive}:
82.8\% (425) of the communities have a {\em Modularity Score} $MS_C \geq 50\%$, which means
that more than half of all possible edges within the community exist.
Interestingly, the largest communities exhibit strong cohesiveness.
In Table~\ref{tab:communitysummary},  we present a high-level profile of the five largest communities which have a  Modularity Score  of $0.65$-$0.78$, which is indicative of tightly-connected communities.

{\it b. Who are the community leaders?}
We want to identify the influential authors as part of profiling a community.
We identify the top two most influential \authoritys and \hubs per community using the \hscore from Section~\ref{sec:influence}. This leads us to a group of 144 leaders of the communities of size of at least 20
authors. We find 81\% of these community leaders are part of the Highly Influential Group (HIG) of authors.
This suggests that the HIG authors are indeed driving forces for the ecosystem.
In the future, we intend to investigate in more depth the influence
and dynamics of each community.





{\it c. What is the focus of each community in terms of platform and malware type?}
A security expert would want to know the main
type of malware (e.g. ransomware) and the target platform (e.g. Linux) of a community.
We use the Repository Keyword Set, $W_r$, information of a repository $r$, as we defined in Section~\ref{sec:data},
and we use it to characterize the community. 

One way to quantify the importance of a keyword for a community
is to measure the number of repositories, for which that keyword appears at least once. 
In detail,
we use the {\bf Strength Of Presence} \textit{(SOP)} metric, which we define as follows.
For a community $C$ with a set of $R$ repositories,
we define $k_i$ to be the number of repositories,  in which keyword $i$ appears in the metadata  {\it $W_r$} for repository $r$ at least once 
 for all  repositories $r \in R$.
We define the $SOP_i$ of keyword $i$ from keyword set $S$ as follows: $SOP_i=\frac{k_i}{\sum_{j\in S} k_j}$.
In Table~\ref{tab:communitysummary}, we show the most dominant keywords from malware types and platforms sets for each community and the related SOP scores.

\begin{figure}[t]
    \centering
    \includegraphics[width=0.8\linewidth, height=2.5cm]{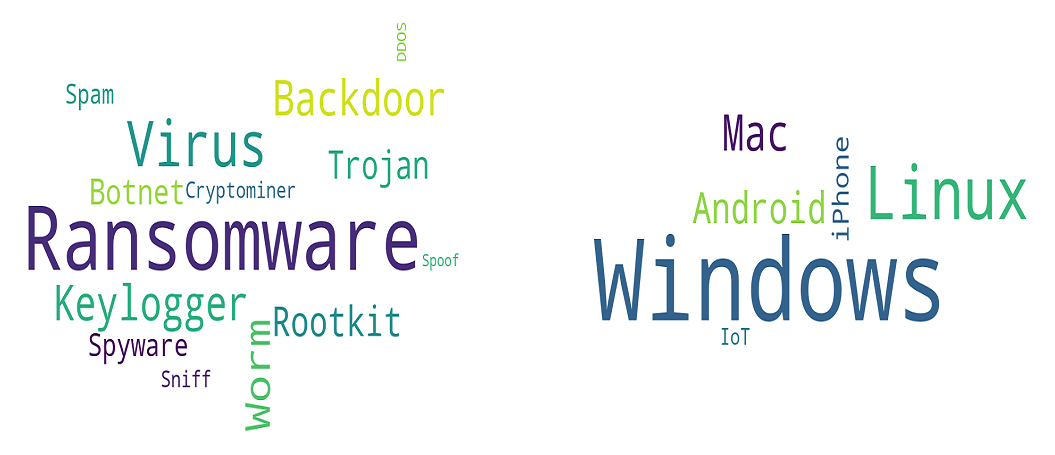}
    \caption{The word-cloud for the malware types and  platforms keywords for the third largest community: Ransomware and Windows dominate.
    }
    \label{fig:worldcloud_community}
\end{figure}

We can also use the {\it SOP}
 to visualize the keywords as a word-cloud.
 A word-cloud is a more immediate, appealing, and visceral
 way to display the information.
In Figure~\ref{fig:worldcloud_community}, 
we show the word-cloud for the third largest community, which is dominated by {\em ransomware} malware and targets {\em Windows} platforms.
Not only we see the main words stand out, but their relative
size conveys their dominance over the other words more viscerally
than a lengthy table of numbers.







We present the results of this type of profiling for the largest communities in Table~\ref{tab:communitysummary}, which
we also discuss below.

\begin{table}[t]
    \centering
    \footnotesize
    \caption{High-level profile   of the five largest  communities of \malauthor and malware repositories.}
    \begin{tabular}
        {|p{0.02\linewidth}|p{0.09\linewidth}|p{0.07\linewidth}|p{0.05\linewidth}||p{0.13\linewidth}|p{0.05\linewidth}||p{0.15\linewidth}|p{0.05\linewidth}|}
        \hline
         \textbf{ID} & \textbf{Authors} & \textbf{Repos} & \textbf{MS}& \textbf{Dominant Platform}& \textbf{SOP}& \textbf{Dominant type}& \textbf{SOP} \\ \hline
         1 & 584 & 677 & 0.65& Linux & 0.32 & Keylogger& 0.29 \\ \hline
         2 & 419 & 544 & 0.67& Windows & 0.26 & Virus & 0.31 \\ \hline
          3 & 175 & 288 & 0.73& Windows & 0.65 & Ransomware & 0.44 \\ \hline
          4 & 57 & 100 & 0.78& Linux & 0.43 & Spyware & 0.43 \\ \hline
          5 & 47 & 57 & 0.71& Mac & 0.33 & Trojan & 0.22 
          \\ \hline
    \end{tabular}
    \label{tab:communitysummary}
\end{table}

We find that the largest community of 584 \malauthor and 677 malware repositories 
having Linux ($SOP=0.32$) and keylogger ($SOP=0.29$) as the dominant platform and malware type.
Interestingly, we  find that 49 of the top 100 most prolific (in terms of the number of repositories created) authors are in this community.
Upon closer investigation,
we find that 11 out of the 15 authors with the highest degree in the subgraph of this community are keylogger developers.

The third-largest community consists of 175 \malauthor and 288 malware repositories and revolves around Ransomware 
($SOP= 0.65$) and Windows platform ($SOP=0.44$). For reference, we present the word-cloud of the malware types and platforms based on the {\em SOP} score in Figure \ref{fig:worldcloud_community} for this community which exhibits that Ransomware and Windows possess the highest $SOP$ scores.

Finally, the fourth largest  community (57 authors, 100 repositories)
is the  most tightly connected ($MS=0.78$) and it  revolves around 
the development of attack tools for Kali Linux. 
Upon closer inspection, we find that 15 of the top 25 authors (based on node degrees) form an approximate bipartite clique with 5 repositories. This group developed {\it WiFiPhisher} in 2016, a Linux-based python 
phishing tool~\cite{wifiphisher},
which has been used for both good and evil~\cite{wifiphisher_malicious}.

The above are indicative of the potential 
information that we could extract from 
these malware repositories.
In the future, we intend to:
(a) extract more detailed textual information from each community, and
(b) study the evolution and dynamics  of these communities over time.





 \section{Author Investigation}
\label{sec:intention}

\begin{figure}[t]
    \centering
    \includegraphics[width=0.65\linewidth]{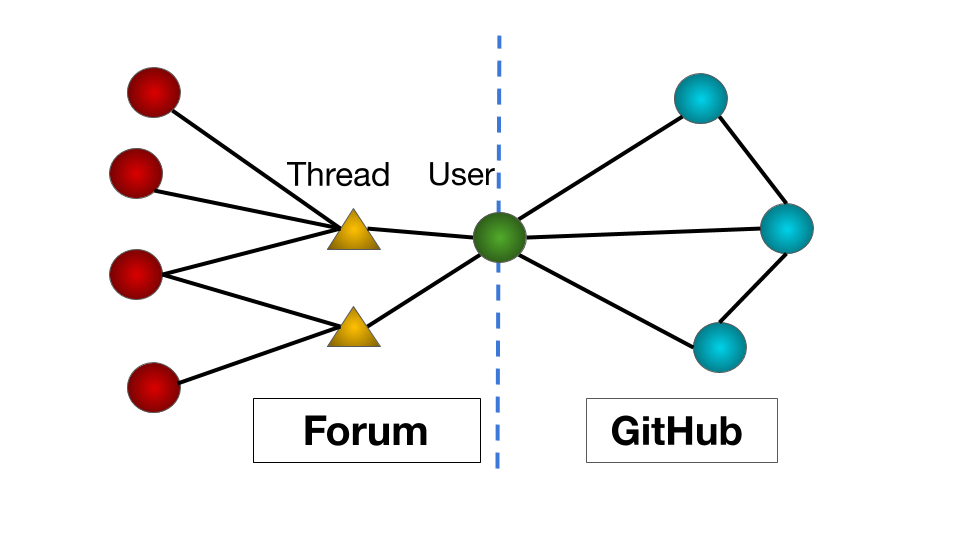}
    \caption{A cross-platform egonet: capturing the neighbors of both the security forum  and \github. 
    }
    \label{fig:cross-egonet}
\end{figure}

\begin{table*}[ht]
\caption{Profiles of four cross-platform users.}
\footnotesize
    \begin{tabular}{|p{0.07\linewidth}||p{0.04\linewidth}|p{0.06\linewidth}|p{0.08\linewidth}||p{0.045\linewidth}|p{0.04\linewidth}|p{0.04\linewidth}|p{0.08\linewidth}|p{0.16\linewidth}|p{0.13\linewidth}|}
    \hline
    {\bf Name} & {\bf Forum} & {\bf Posts in forum} & {\bf Collab/tors in forums}& {\bf Malw. repos} & {\bf Follow-ers} & {\bf Forks} & {\bf Collab/tors in \github} & {\bf Repository content} & {\bf Internet-wide Reputation}\\
    \hline
     misterch0c & WS & 7 & 224 & 7 & 749 & 81 &898 & Cracked malware code & Self-declared hacker\\
    \hline
    3vilp4wn & HTS & 103 & 513 & 1& 0 &1 & 6 & Python keylogger & Keylogger developer\\
    \hline
    fahimmagsi & OC & 73 & 175 & 1 & 1 & 0 & 1& Backdoor &  Famous hacker\\
    \hline
    Evilcry & EH & 18 & 444 & 2 & 89 & 15 & 98& Botnet and ransomware & Ransomware expert \\
    \hline
    
    \end{tabular}
    \label{tab:intention}
\end{table*}


{\it ``Who are these malware authors?''}
To answer this question, we go across platforms
to security forums and
leverage our datasets from several security forums.
 The functions described here are part of the {\em author investigation} module of Figure ~\ref{fig:overview}.

{\bf a. \MalAuthors strive for an online ``brand'' and usernames seem persistent across online platforms.}
We find that many \malauthor 
use the same username consistently across many online platforms, such as security forums, possibly in pursuit of a reputation.



We identify \malHackersNo \malauthor who are active
in  one of our four 
security forums:  12 in Wilders Security, 
6 in Ethical Hacker, 4 in Offensive Community, and 8 in Hack This Site ~\cite{secforums}. 
We argue that some of these usernames correspond to the same users based on the following two observations.

First, we find significant overlap in the interests of the cross-platform usernames. For example,
 usernames {\em int3grate} and {\em jedisct1} 
show interest in ransomware in both platforms, 
while {\em 3vilp4wn} advertises her keylogger malware 
(github.com/ 3vilp4wn/CryptLog) in the 
forum.
Second, these usernames are fairly uncommon, which 
increases the likelihood of belonging to the same person.
For example, the top ten results from internet searching for the username of  author {\em Misterch0c} returns nine  hacker related sites
and a twitter account with a different handle 
but  claimed by Misterch0c.

Note that not all the malware authors or repositories have a malicious purpose. For instance, the project ``Empire''~\cite{project_empire} by {\em xorrior} was created as an offensive tool to stress-test
the security of systems.
However,  it has recently been used by the state-sponsored hacking group \textit{Deep Panda}~\cite{state_sponsored_hacking}. 
In general,  offensive security tools contribute to the power of the malware ecosystem
irrespective of the intention of its creator.



{\bf b. Modeling the cross-platform interactions.}
We propose to study the cross-platform interactions
between \github and security forums as a promising research direction that can bridge two domains: software repositories and online forums. 

We define the {\bf cross-platform egonet} of a user as one that consists of her egonets from the two platforms as shown in Figure ~\ref{fig:cross-egonet}.
The forum egonet  captures the interaction of the users that post on the same threads, while we leverage the Author-Author network to define the \github egonet.

{\bf The value of cross-platform analysis.}
Using the cross-platform egonet as a basis,
we can model the cross-platforms user dynamics, and
more specifically, we can:
(a) identify common ``friends" between the ego-nets,
(b) find the topics of interest and activities in each egonet,
and
(c) model information flow and influences across platforms.
In Figure~\ref{fig:scree-egonet}, we visualize the activity of a cross-platform user by comparing the number of users on each side of the egonet as shown in Figure~\ref{fig:cross-egonet}. 
 In Table~\ref{tab:intention}, we show the actual values of indicative users,
 including the three outliers in the plot.

The cross-platform egonet analysis can  enrich the profile of each user significantly. For example, if we were just looking at \github, we may not have paid attention to {\it 3vilp4wn} and {\it Evilcry}. Both of these authors are less active on \github (small \github egonet), but are quite active in the security forums (large forum egonet).
A closer investigation of the security forums reveals activities that match their interests on \github. 
This suggests that their \github activity is part of their online brand.
For example, {\it 3vilp4wn} advertises her \github keylogger repository in the forum.
We intend to expand in this promising  direction in the future.





{\bf c. Using information from the web.} In our approach, we leverage existing information on hackers from (a) security outlets and databases, and (b) using web queries. With our python-based query and analysis tools, we  verified  the role and activities of authors,
which we omit due to space limitations.

\section{Related Works}
\label{sec:related}


Studying the dynamics of the malware ecosystem on GitHub
has received very little attention.
Most studies differ from our work in that:
(a) they do not focus on malware on \github,
and (b) when they do, they do not take an author-centric angle
as we do here:
they focus on classifying malware repositories
or  use a small set for a particular 
research study.

Our work  builds on our earlier effort~\cite{rokon2020source}, whose main goal is to identify  malware repositories on \github at scale,
but it does not study the malware author ecosystem as we do here.

 


{\bf a. Studies of malware repositories on \github:}
Several other efforts have manually collected a small number of
malware repositories with the purposes of a research study~\cite{lepik2018art,zhong2015stealthy}.
Some other studies~\cite{calleja2016look}~\cite{calleja2018malsource}
analyze malware source code from a software engineering perspective,
but use only a small number 
of \github repositories as a reference. 

{\bf b. Studies of benign repositories on \github:} 
Many studies analyze benign repositories on \github 
from a point of view of software engineering or as a social network. 
Some efforts find influential users and analyze the motivation behind following, forking, and contributions~\cite{blincoe2016understanding, jiang2017and}.
Earlier efforts study repositories by analyzing 
the repository-repository relationship graph~\cite{thung2013network}, and
by using an activity graph \cite{xavier2014understanding}.


Several works in this area
 identify influential authors
and repositories using: the starring activity~\cite{hu2016influence}, the Following-Star-Fork activity~\cite{hu2018user},  or a rank-based approach~\cite{liao2017devrank}. 
Note that  a version of the hyperlink-induced topic search algorithm~\cite{li2002improvement} has been used
by some of the above efforts for calculating influence,
but they do not adjust the weights to account for the different frequencies of the types of interactions between users.

For our bipartite clustering, we adapt
the greedy modularity maximization approach~\cite{clauset2004finding}\cite{alzahrani2016community}.


{\bf c. Studies on security forums:} 
This is a recent and less studied area of research. Most of the works focus on  extracting entities of interest in security forums.
An interesting study focuses on the dynamics of the black-market of hacking goods and services and their pricing~\cite{portnoff2017tools}. Other studies focus on identifying important events and threats \cite{Sapienza2017_USC1, Sapienza2018_USC2}.
Several studies focus on identifying key actors in security forums by utilizing their social and linguistics behavior \cite{Marin2018_keyhacker}. None of the aforementioned works focus on the dynamics among hackers across platforms.

{\bf d. Cross-platform study:} Finally, some efforts 
study author activities on different
software development forums, namely \github and Stack Overflow~\cite{hauff2015matching, lee2017github}, but do
not consider information from security forums. 


\section{Conclusion}
\label{sec:conclusion}

We develop a systematic approach for studying the ecosystem of hackers. Our approach develops methods to identify (a) influential hackers, (b) communities of collaborating hackers, and (c) their cross-platform interactions.
Our study concludes in three key takeaway messages:
(a) the malware ecosystem is substantial and growing rapidly,
(b) it is highly collaborative,
and (c) it contains  professional malicious hackers. 


Our initial findings are just the beginning of a 
promising future effort that can shed light on this
online malware author ecosystem, which spans software repositories
and security forums.
The current work thus can be seen as a building block that can enable
new research directions.


Follow up research can expand on our work to develop preemptive security initiatives, such as: (a) monitoring hacker activity, (b) detecting emerging trends, and (c) 
identifying particularly influential hackers towards safeguarding the Internet.

\section{Acknowledgement}
This work was supported by the UC Multicampus-National Lab Collaborative Research and Training (UCNLCRT) award \#LFR18548554.

{ \bibliographystyle{IEEEtran}}
\bibliography{ROKON}

\begin{thebibliography}{10}
\providecommand{\url}[1]{#1}
\csname url@samestyle\endcsname
\providecommand{\newblock}{\relax}
\providecommand{\bibinfo}[2]{#2}
\providecommand{\BIBentrySTDinterwordspacing}{\spaceskip=0pt\relax}
\providecommand{\BIBentryALTinterwordstretchfactor}{4}
\providecommand{\BIBentryALTinterwordspacing}{\spaceskip=\fontdimen2\font plus
\BIBentryALTinterwordstretchfactor\fontdimen3\font minus
  \fontdimen4\font\relax}
\providecommand{\BIBforeignlanguage}[2]{{%
\expandafter\ifx\csname l@#1\endcsname\relax
\typeout{** WARNING: IEEEtran.bst: No hyphenation pattern has been}%
\typeout{** loaded for the language `#1'. Using the pattern for}%
\typeout{** the default language instead.}%
\else
\language=\csname l@#1\endcsname
\fi
#2}}
\providecommand{\BIBdecl}{\relax}
\BIBdecl

\bibitem{17yearsoldboy}
{Aaron Holmes}, ``17 years old boy tried to hack twitter,''
  \url{https://www.businessinsider.com/twitter-hacker-florida-teen-past-minecraft-bitcoin-scams-2020-8/},
  August 2020.

\bibitem{rokon2020source}
M.~O.~F. Rokon, R.~Islam, A.~Darki, E.~E. Papalexakis, and M.~Faloutsos,
  ``Sourcefinder: Finding malware source-code from publicly available
  repositories,'' in \emph{RAID 2020 23rd International Symposium on Research
  in Attacks, Intrusions and Defenses}, 2020.

\bibitem{Joobin}
J.~Gharibshah, E.~E. Papalexakis, and M.~Faloutsos, ``{REST}: A thread
  embedding approach for identifying and classifying user-specified information
  in security forums.'' \emph{ICWSM}, 2020.

\bibitem{portnoff2017tools}
R.~S. Portnoff, S.~Afroz, G.~Durrett, J.~K. Kummerfeld, T.~Berg-Kirkpatrick,
  D.~McCoy, K.~Levchenko, and V.~Paxson, ``Tools for automated analysis of
  cybercriminal markets,'' in \emph{WWW}, 2017, p. 657.

\bibitem{calleja2016look}
A.~Calleja, J.~Tapiador, and J.~Caballero, ``A look into 30 years of malware
  development from a software metrics perspective,'' in \emph{International
  Symposium on Research in Attacks, Intrusions, and Defenses}.\hskip 1em plus
  0.5em minus 0.4em\relax Springer, 2016, pp. 325--345.

\bibitem{calleja2018malsource}
A.~Calleja, J.~Tapiador, and J.~Cabalero, ``The malsource dataset: Quantifying
  complexity and code reuse in malware development,'' \emph{IEEE Transactions
  on Information Forensics and Security}, vol.~14, no.~12, pp. 3175--3190,
  2018.

\bibitem{blincoe2016understanding}
K.~Blincoe, J.~Sheoran, S.~Goggins, E.~Petakovic, and D.~Damian,
  ``Understanding the popular users: Following, affiliation influence and
  leadership on github,'' \emph{Information and Software Technology}, vol.~70,
  pp. 30--39, 2016.

\bibitem{Sapienza2017_USC1}
A.~Sapienza, A.~Bessi, S.~Damodaran, P.~Shakarian, K.~Lerman, and E.~Ferrara,
  ``Early warnings of cyber threats in online discussions,'' in \emph{2017 IEEE
  International Conference on Data Mining Workshops (ICDMW)}, Nov 2017, pp.
  667--674.

\bibitem{Sapienza2018_USC2}
A.~Sapienza, S.~K. Ernala, A.~Bessi, K.~Lerman, and E.~Ferrara, ``Discover:
  Mining online chatter for emerging cyber threats,'' in \emph{Companion
  Proceedings of the The Web Conference 2018}.\hskip 1em plus 0.5em minus
  0.4em\relax International World Wide Web Conferences Steering Committee, pp.
  983--990.

\bibitem{secforums}
{Security Forums}, ``Ethical hacker, hack this site, offensive community,
  wilders security,'' \url{https://www.ethicalhacker.net/},
  \url{https://www.hackthissite.org/}, \url{http://offensivecommunity.net/},
  \url{https://www.wilderssecurity.com/}.

\bibitem{jiang2017and}
J.~Jiang, D.~Lo, J.~He, X.~Xia, P.~S. Kochhar, and L.~Zhang, ``Why and how
  developers fork what from whom in github,'' \emph{Empirical Software
  Engineering}, vol.~22, no.~1, pp. 547--578, 2017.

\bibitem{weng2010twitterrank}
J.~Weng, E.-P. Lim, J.~Jiang, and Q.~He, ``Twitterrank: finding topic-sensitive
  influential twitterers,'' in \emph{Proceedings of the third ACM international
  conference on Web search and data mining}, 2010, pp. 261--270.

\bibitem{wifiphisher}
{Sophron}, ``Wifiphisher,'' \url{https://github.com/wifiphisher/wifiphisher},
  2014, [Online; accessed 14-March-2020].

\bibitem{wifiphisher_malicious}
{Cybersec}, ``Stealing password in 5 minutes using wifiphisher,''
  \url{https://www.secjuice.com/phishing-with-wifiphisher/}, 2018.

\bibitem{project_empire}
{EmpireProject}, ``Project empire,''
  \url{https://github.com/EmpireProject/Empire}.

\bibitem{state_sponsored_hacking}
{Mitre}, ``State sponsored hacking tool,''
  \url{https://attack.mitre.org/software/S0363}, 2019.

\bibitem{lepik2018art}
T.~Lepik, K.~Maennel, M.~Ernits, and O.~Maennel, ``Art and automation of
  teaching malware reverse engineering,'' in \emph{International Conference on
  Learning and Collaboration Technologies}.\hskip 1em plus 0.5em minus
  0.4em\relax Springer, 2018, pp. 461--472.

\bibitem{zhong2015stealthy}
X.~Zhong, Y.~Fu, L.~Yu, R.~Brooks, and G.~K. Venayagamoorthy, ``Stealthy
  malware traffic-not as innocent as it looks,'' in \emph{2015 10th
  International Conference on Malicious and Unwanted Software}.\hskip 1em plus
  0.5em minus 0.4em\relax IEEE, 2015, pp. 110--116.

\bibitem{thung2013network}
F.~Thung, T.~F. Bissyande, D.~Lo, and L.~Jiang, ``Network structure of social
  coding in github,'' in \emph{2013 17th European conference on software
  maintenance and reengineering}.\hskip 1em plus 0.5em minus 0.4em\relax IEEE,
  2013, pp. 323--326.

\bibitem{xavier2014understanding}
J.~Xavier, A.~Macedo, and M.~de~Almeida~Maia, ``Understanding the popularity of
  reporters and assignees in the github.'' in \emph{SEKE}, 2014.

\bibitem{hu2016influence}
Y.~Hu, J.~Zhang, X.~Bai, S.~Yu, and Z.~Yang, ``Influence analysis of github
  repositories,'' \emph{SpringerPlus}, vol.~5, no.~1, pp. 1--19, 2016.

\bibitem{hu2018user}
Y.~Hu, S.~Wang, Y.~Ren, and K.-K.~R. Choo, ``User influence analysis for github
  developer social networks,'' \emph{Expert Systems with Applications}, vol.
  108, pp. 108--118, 2018.

\bibitem{liao2017devrank}
Z.~Liao, H.~Jin, Y.~Li, B.~Zhao, J.~Wu, and S.~Liu, ``Devrank: Mining
  influential developers in github,'' in \emph{GLOBECOM 2017-2017 IEEE Global
  Communications Conference}.\hskip 1em plus 0.5em minus 0.4em\relax IEEE,
  2017, pp. 1--6.

\bibitem{li2002improvement}
L.~Li, Y.~Shang, and W.~Zhang, ``Improvement of hits-based algorithms on web
  documents,'' in \emph{Proceedings of the 11th international conference on
  World Wide Web}, 2002, pp. 527--535.

\bibitem{clauset2004finding}
A.~Clauset, M.~E. Newman, and C.~Moore, ``Finding community structure in very
  large networks,'' \emph{Physical review E}, vol.~70, no.~6, p.~6, 2004.

\bibitem{alzahrani2016community}
T.~Alzahrani and K.~J. Horadam, ``Community detection in bipartite networks:
  Algorithms and case studies,'' in \emph{Complex systems and networks}.\hskip
  1em plus 0.5em minus 0.4em\relax Springer, 2016, pp. 25--50.

\bibitem{Marin2018_keyhacker}
E.~Marin, J.~Shakarian, and P.~Shakarian, ``Mining key-hackers on darkweb
  forums,'' in \emph{2018 1st International Conference on Data Intelligence and
  Security (ICDIS)}, April 2018, pp. 73--80.

\bibitem{hauff2015matching}
C.~Hauff and G.~Gousios, ``Matching github developer profiles to job
  advertisements,'' in \emph{2015 IEEE/ACM 12th Working Conference on Mining
  Software Repositories}.\hskip 1em plus 0.5em minus 0.4em\relax IEEE, 2015,
  pp. 362--366.

\bibitem{lee2017github}
R.~K.-W. Lee and D.~Lo, ``Github and stack overflow: Analyzing developer
  interests across multiple social collaborative platforms,'' in
  \emph{International Conference on Social Informatics}.\hskip 1em plus 0.5em
  minus 0.4em\relax Springer, 2017, pp. 245--256.

\end{thebibliography}

\end{document}